\documentclass[aps,prl,twocolumn,showpacs,preprintnumbers,amsmath,amssymb]{revtex4-1}

\usepackage{graphicx}
\usepackage{bm}
\usepackage{upgreek}

\begin{document}


\title{Observation of the dominant spin-triplet supercurrent in Josephson spin valves with strong Ni ferromagnets}

\author{O. M. Kapran$^1$ , A. Iovan$^1$, T. Golod$^1$ and V. M. Krasnov$^{1,2}$ }
\email[E-mail: ]{Vladimir.Krasnov@fysik.su.se}


\affiliation{$^1$ Department of Physics, Stockholm University,
AlbaNova University Center, SE-10691 Stockholm, Sweden}
\affiliation{$^2$ Moscow Institute of Physics and Technology,
State University, 9 Institutskiy per., Dolgoprudny, Moscow Region
141700 Russia}

\begin{abstract}
We study experimentally nanoscale Josephson junctions and
Josephson spin-valves containing strong Ni ferromagnets. We
observe that in contrast to junctions, spin valves with the same
geometry exhibit anomalous $I_c(H)$ patterns with two peaks
separated by a dip. We develop several techniques for {\em
in-situ} characterization of micromagnetic states in our
nano-devices, including magnetoresistance, absolute Josephson
fluxometry and First-Order-Reversal-Curves analysis. They reveal a
clear correlation of the dip in supercurrent with the antiparallel
state of a spin-valve and the peaks with two noncollinear magnetic
states, thus providing evidence for generation of spin-triplet
superconductivity. A quantitative analysis brings us to a
conclusion that the triplet current in out Ni-based spin-valves is
approximately three times larger than the conventional singlet
supercurrent.

\end{abstract}



\maketitle

\section{I. Introduction}

Spin-polarized ferromagnetism is antipathetic to spin-singlet
superconductivity. However, destruction of singlet Cooper pairs in a ferromagnet is
not an instant process. Pairing correlations survive over a
certain time/distance, during which precession of spins in exchange field may create a correlated triplet pair. The
corresponding odd-frequency spin-triplet order parameter has been
predicted theoretically using various approaches
\cite{Buzdin2005,Efetov2005,Fominov,Blanter2004,Eschrig,Houzet_2007,Golubov,Trifunovic_2011,Trifunovic_PRL2011,Melnikov_2012,Pugach_2012,Linder_2012,Richard_2015,Hikino_2015,Linder_2015,Ren_2016}.
This inspired intense experimental search for this exotic state in
Superconductor/Ferromagnet (S/F) heterostructures
\cite{Bell_2004,Robinson_2010,Iovan_2014,Dresselhaus_2014,Ovsyannikov,Robinson_Sc2010,Khaire_2010,Banerjee_2014,Glick_2017,Aarts_2017,Martinez_2016,Strunk_2017,Sidorenko_2016}.
Although supercurrents through F were reported many
times \cite{Bell_2004,Robinson_2010,Iovan_2014,Dresselhaus_2014,Ovsyannikov,Robinson_Sc2010,Khaire_2010,Banerjee_2014,Glick_2017,Aarts_2017,Martinez_2016,Strunk_2017},
it is difficult to prove their triplet nature. First, even
spin-singlet current can flow over long ranges 
in clean or weak ferromagnets \cite{Buzdin2005,Eschrig,Oboznov_2006,Melnikov_2012}. The singlet
current is reduced in strong F, which should be material of choice for a critical test. Second, the supercurrent
strongly depends on usually unknown domain structure in F
\cite{Iovan_2014,Aarts_2017,Weides_2008}, flux quantization in
S \cite{Golovchanskiy_2016,Iovan_2017}, both influenced by size and geometry. This uncertainty can be obviated in
nanoscale devices with mono- (or few) domain F-layers and with the
flux-quantization field larger than the coercive field
\cite{Iovan_2017}. Finally, the long-range triplet current should
appear only in the noncollinear magnetic state
\cite{Efetov2005,Houzet_2007,Golubov,Trifunovic_2011,Trifunovic_PRL2011,Melnikov_2012}.
Therefore, unambiguous identification of the pairing order is
only possible if the micromagnetic state of the actual device is
known. It is not sufficient to analyze similar large-area
heterostructures because their magnetic properties (coercive
fields, domain structure, shape anisotropy) would be different
from a nano-device. In order to prove/disprove the triplet nature
of supercurrent it is necessary to demonstrate its
correlation/anticorrelation with the noncollinear state
\cite{Iovan_2014,Dresselhaus_2014,Ovsyannikov,Banerjee_2014,Glick_2017,Aarts_2017,Martinez_2016,Strunk_2017}.
In the end, it is all about having an {\em in-situ} control over the
micromagnetic state of the studied nano-device. This is our main
motivation.

The noncollinear magnetic state can be controllably created in
mono-domain spin valves. The simplest is the pseudo spin valve
F$_1$NF$_2$ with two F$_{1,2}$ layers separated by a normal metal
(N) spacer. Triplet current in this case is second-harmonic with
respect to the phase difference and is proportional to the difference between F$_1$ and F$_2$
\cite{Trifunovic_2011,Trifunovic_PRL2011,Melnikov_2012,Richard_2015,Hikino_2015,Ren_2016} (see the Appendix for more details). Therefore, an asymmetric spin-valve F$_1 \neq$F$_2$ is needed for
generation of the triplet current. The asymmetry (different coercive fields) is also needed for controllable tuning of the
relative magnetization angle $\alpha$ between F$_{1,2}$-layers.

Here we study experimentally nano-scale SFS Josephson junctions (JJ's) and
SF$_1$NF$_2$S Josephson spin-valves (JSV's).  We use strong
F (Ni) to suppress singlet currents and to make triplet currents dominant. We focus on development of various methods for
{\em in-situ} characterization of micromagnetic states in our nano-devices, including
magnetoresistance (MR), absolute Josephson fluxometry (AJF) and First-Order-Reversal-Curves (FORC) analysis.
We observe that JSV's behave qualitatively differently compared to JJ's with
similar geometry: they exhibit non-Fraunhofer $I_c(H)$ modulation with two distinct
peaks separated by a dip. {\em In-situ} characterization reveals a clear correlation of the supercurrent dip with the antiparallel (AP) state of the JSV and the peaks with two noncollinear
states arround it. This provides an {\em in-situ} evidence for generation of spin-triplet superconductivity.

\begin{figure}[t]
    \centering
    \includegraphics[width=0.4\textwidth]{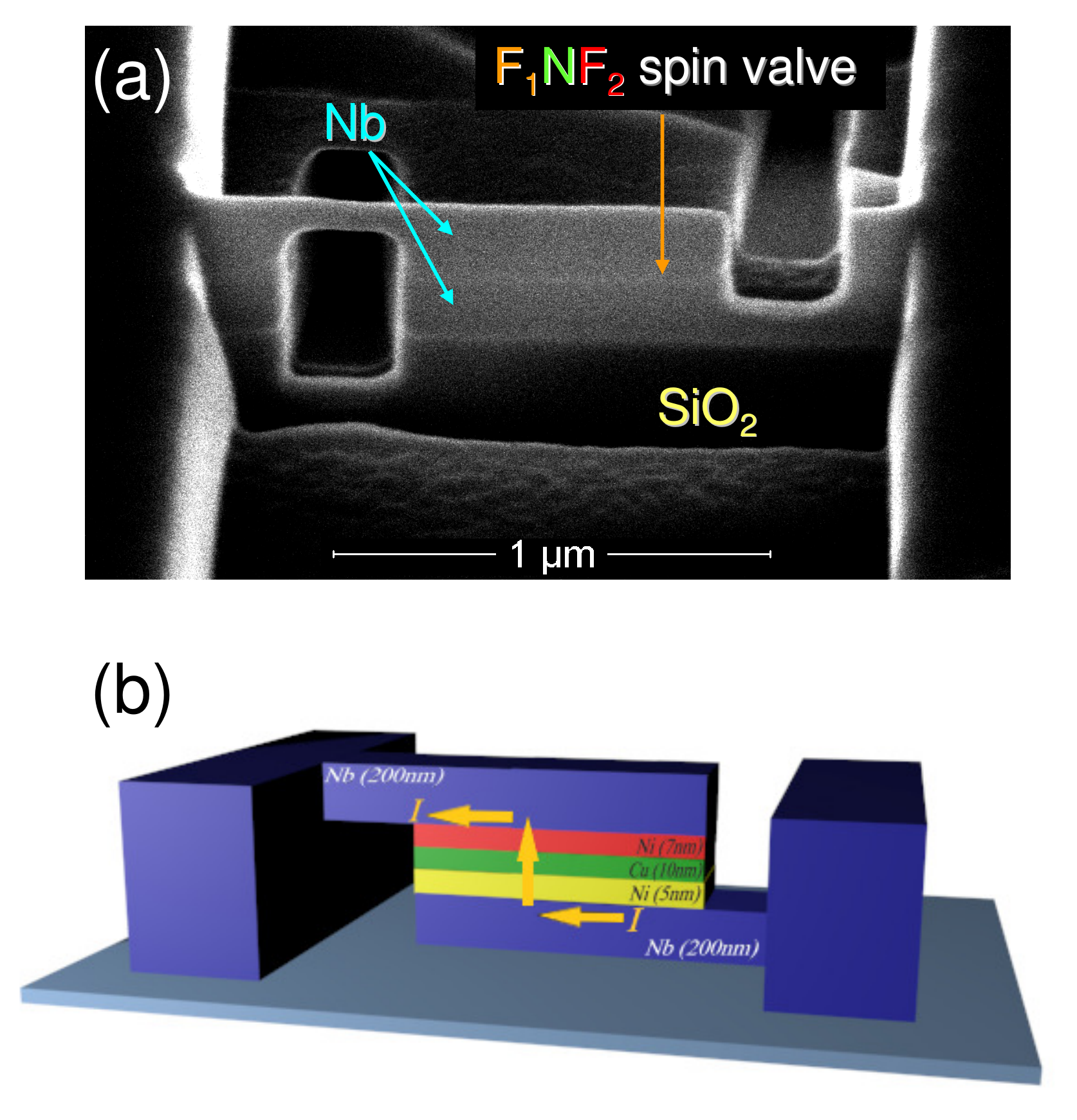}
    \caption{(color online). (a) SEM image of one of studied Josephson spin valves.
    (b) A clarifying sketch of the device. Arrows indicate the
    zig-zag current flow path. The transport current is sent along the bridge, but, because of the two FIB cuts, it is forsed to flow accross F-layers.
}
    \label{fig:fig1}
\end{figure}

\section{II. Experimental}

We study nanoscale Nb(200nm)/Ni(5nm)-Cu(10nm)-Ni(7nm)/Nb(200nm)
JSV's and Nb(200nm)/Ni(5-10nm)/Nb(200nm) JJ's. Thin film multilayers are deposited by dc-magnetron sputtering
and patterned into $\mu$m size bridges by photolithography and reactive ion
etching. Subsequently they are transferred into a dual-beam Scanning Electron Microscope (SEM) / Focused Ion Beam (FIB) and
nanoscale devices are made by 3D FIB nanosculpturing
\cite{Golod_2010,Iovan_2014}. Both JJ's and JSV's have similar rectangular shapes with short sides
$100-300$ nm and long sides $250-1400$ nm.
Several devices with different sizes are made at the same chip. Figure \ref{fig:fig1} shows a
SEM image of one of the studied JSV's with a clarifying sketch.

Measurements are performed in closed-cycle cryostats.
For analysis of $I_c(H)$ modulation the in-plane magnetic field is applied either
parallel ($H_{\parallel}$, along the easy magnetization axis) or perpendicular ($H_{\perp}$, along the hard axis) to the long side.
More details can be found in the Supplementary \cite{Supplem}.

\begin{figure*}[t]
    \centering
    \includegraphics[width=0.9\textwidth]{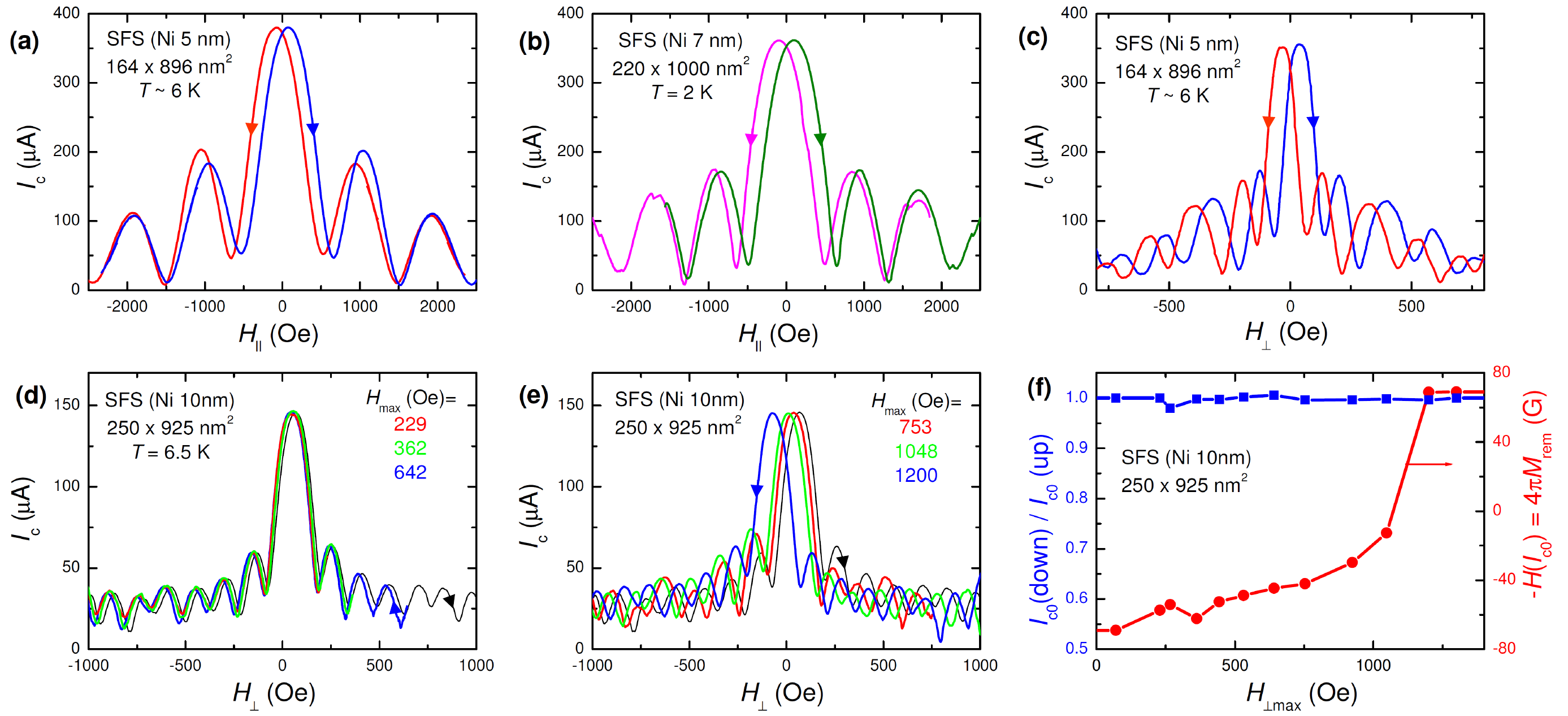}
    \caption{(color online). Characteristics of SFS (Nb/Ni/Nb) Josephson junctions with different Ni thicknesses.
    (a) and (b) $I_c(H_{\parallel})$ patterns for up and down field sweeps for junctions with (a) 5 nm and (b) 7 nm Ni thickness
    in field parallel to the long side. (c) $I_c(H_{\perp})$ modulation for the same Ni (5 nm) junction in the field perpendicular
    to the long side. Note that $I_c(H)$ has Fraunhofer-type modulation at both field orientations.
    (d-f) First-order-reversal-curves analysis of $I_c(H_{\perp})$ patterns for a junction with Ni (10 nm). Black curves in (d,e) represent the upwards sweep from the saturated negative state.
    Red, green and blue curves represent reversal curves with different $H_{max}$. (f) The summary of FORC analysis from (d,e). Red circles (right axis) show the position of the central maximum, representing the remnant magnetization in the junction. Blue squares (left axis) show the
    amplitude of the central peak. It is seen that remagnetization of the SFS junction leads to a
trivial hysteresis: the $I_c(H)$ patterns maintain their shapes
and merely shift due to changing magnetization of the F-interlayer.}
    \label{fig:fig2}
\end{figure*}

\section{III. Properties of SFS junctions}

Figures \ref{fig:fig2} (a,b) represent measured $I_c(H_{\parallel})$
patterns in the easy axis orientation for JJ's (a) Ni (5 nm) with area $164\times 896$ nm$^2$, and (b) Ni (7 nm) with
areas $200\times 1000$ nm$^2$. Fig. \ref{fig:fig2} (c) shows
$I_c(H_{\perp})$ along the hard axis for the same Ni (5 nm) JJ. Up and down field sweeps are shown. They exhibit
hysteresis due to finite coercivity. From Figs. \ref{fig:fig2}
(a,b) it can be seen that the hysteresis starts/ends at about
$\sim \pm 1.5$ kOe, which represents the saturation field.
In all cases SFS JJ's exhibit regular Fraunhofer-type $I_c(H)$, indicating good uniformity of Ni-interlayers \cite{Krasnov_1997}.

\subsection{A. First-Order-Reversal-Curves analysis}

FORC is a powerful tool for {\em in-situ}
characterization of magnetic states in complex ferromagnetic
structures \cite{FORC1_2008,FORC2_2013,Akerman_2014}. The analysis starts at the
same saturated state. Then field is swept to a reversal field
$H_{max}$ and measurements are carried out on the
way back to the saturated state.
Figs. \ref{fig:fig2} (d,e) represent $I_c(H_{\perp})$ FORC's for a JJ with
Ni (10 nm). Thin black lines represent the upward
sweep from the saturated $\downarrow\downarrow$ state. Red, green and blue lines are FORCs with different
$H_{\perp max}$. FORC's show very little hysteresis up to $H_{\perp max}\sim 1.1$ kOe and then
rapidly jump to the saturated $\uparrow\uparrow$ state. This reflects an
abrupt remagnetization of the Ni-nanoparticle within the JJ. Note, that the curves for different $H_{\perp max}$ in
Figs. \ref{fig:fig2} (d,e) have the same Fraunhofer-type
shape, which just shifts upon remagnetization of the Ni-interlayer.

\subsection{B. Absolute Josephson fluxometry}

AJF is based on flux quantization in Josephson devices, due to which minima
and maxima of $I_c$ occur at integer and half-integer flux quanta $\Phi_0$ within a device. Magnetization is related to flux via: $M=(\Phi/L d^* -H)/4\pi$, where $L$ is the
size and $d^*$ the magnetic thickness of the device. Thus absolute values of $M$ can be obtained at discrete fields determined by the flux quantization field $\Delta H =\Phi_0/Ld^*$ \cite{Bolginov_2012,Iovan_2014}.

\subsection{C. Combined AJF+FORC}

For nano-devices with large $\Delta H$ the discretness of AJF is a limitation. To obviate this problem we combine AJF with FORC, which allows continuous determination of
$M(H_{max})$ for arbitrary small devices. For example, the central maxima of FORC's in Figs. \ref{fig:fig2} (d,e) correspond to $\Phi=0$. Therefore, fields at which they occur, $H(I_{c0})$, represent absolute values of
remnant magnetization $M_{rem} = -H(I_{c0})/4\pi$. Since we can vary $H_{max}$ with arbitrary small step, we can get a continuous
$M_{rem}(H_{max})$ curve from such AJF+FORC analysis even for very small devices. This is demonstrated in Fig. \ref{fig:fig2} (f) where red circles represent
$-H(I_{c0})=M_{rem}/4\pi$ as a function of $H_{\perp max}$ for FORC's from Fig. \ref{fig:fig2} (d,e).  It is seen that $M_{rem}$ switches rapidly at $H_{\perp max}\gtrsim 1.1$ kOe, which
represents the coercive field.
Blue squares represent $I_{c0}$, which apparently stays constant. Thus, hysteresis in SFS JJ's is {\em trivial}:
remagnetization of the F-layer changes the internal flux, which just shifts $I_c(H)$ patterns without changing their shapes.

\section{IV. Properties of Josephson spin valves}

\begin{figure*}[t]
    \centering
    \includegraphics[width=0.9\textwidth]{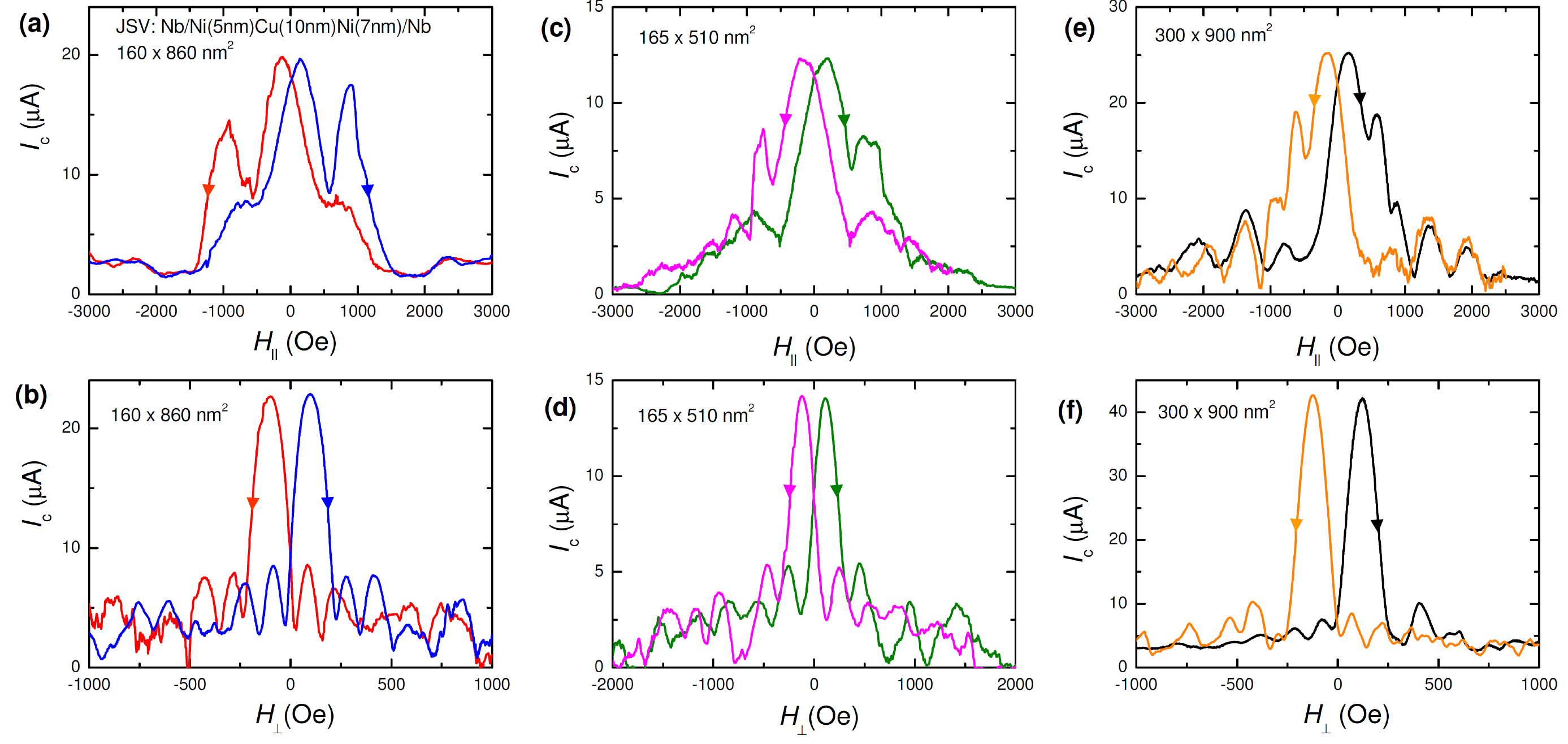}
    \caption{(color online). $I_c(H)$ modulation patterns for three Josephson spin valves with different sizes, made at the same chip. (a,c,e) in fields parallel and (b,d,f) perpendicular to the long side.
(a,b) JSV $160 \times 860$ nm$^2$ at
    $T\simeq 2$ K, (c,d) JSV $165 \times 510$ nm$^2$ at $T\simeq 2$ K. (e,f) JSV $300 \times 900$ nm$^2$ at $T\simeq 0.6$ K.
    Note non-Fraunhofer double-peak patterns in parallel field orientation.}
    \label{fig:fig3}
\end{figure*}

Figure \ref{fig:fig3} shows $I_c(H)$ patterns for three JSV's with
different sizes from the same chip in easy (a,c,e) and hard (b,d,f) axis
orientations. In a stark contrast to SFS JJ's, Figs. \ref{fig:fig2} (a,b), JSV's with a similar geometry exhibit a profound
distortion of the central $I_c(H)$ maximum in the easy axis
orientation. The distortion depends on the size. For the
narrowest JSV's, (a) $L=160$ nm, the central
maximum splits into two peaks, separated by a dip. With increasing
JSV size, (e) $L=300$ nm, the splitting decreases. For the hard axis orientation, corresponding
to larger sizes: (d) 510 nm, (b) 860 nm and (f) 900 nm, the distortion seemingly disappears and $I_c(H)$ patterns acquire Fraunhofer-type shapes. The latter indicates good uniformity of the barrier
\cite{Krasnov_1997}. Therefore, the double-peak distortion in the
easy axis orientation {\em for the same devices} can not be
ascribed to non-uniformity or defects. This is our central
observation that we will analyze below.

\subsection{A. Hard-axis orientation}

We start with the hard-axis orientation
because in this case $I_c(H_{\perp})$ patterns have Fraunhofer-type shapes
facilitating similar analysis as for SFS junctions. Figure
\ref{fig:fig4} (a) represents $I_c(H_{\perp})$ FORC's for a JSV
$300\times900$ nm$^2$. Thin white
lines represent the upward sweep and thick color lines the
FORC's with different $H_{\perp max}$. The lower curve indicates that up to the end
of the central peak, $H_{\perp max}\lesssim 274$ Oe, FORC's are fully reversible. Above it a hysteresis appears.
However, in contrast to SFS JJ's, see Fig. \ref{fig:fig2}
(e), the hysteresis in {\em non-trivial}: Remagnetization of JSV's leads {\em both} to the shift and distortion of $I_c(H)$ patterns. In particular, it leads to a significant reduction of the central maximum, $I_{c0}$, which reaches minimum at $H_{\perp max}\simeq 718$ Oe. With further increase of $H_{\perp max}$, $I_{c0}$ grows back and recovers to the original value when $H_{\perp max}$ exceed the saturation field, see Fig.
\ref{fig:fig4} (b). Figs. \ref{fig:fig4} (c,d) represent AJF+FORC analysis: (c) $M_{rem}(H_{\perp max})$ and (d) $I_{c0}(H_{\perp max})$. It is seen that $I_{c0}$ is reduced by up to a factor two
within the hysteresis region, marked by vertical lines, demonstrating the non-trivial type of hysteresis, compared to SFS junctions, see Fig. \ref{fig:fig2} (f).

\subsection{B. Easy-axis orientation}

\begin{figure}[t]
    \centering
    \includegraphics[width=0.5\textwidth]{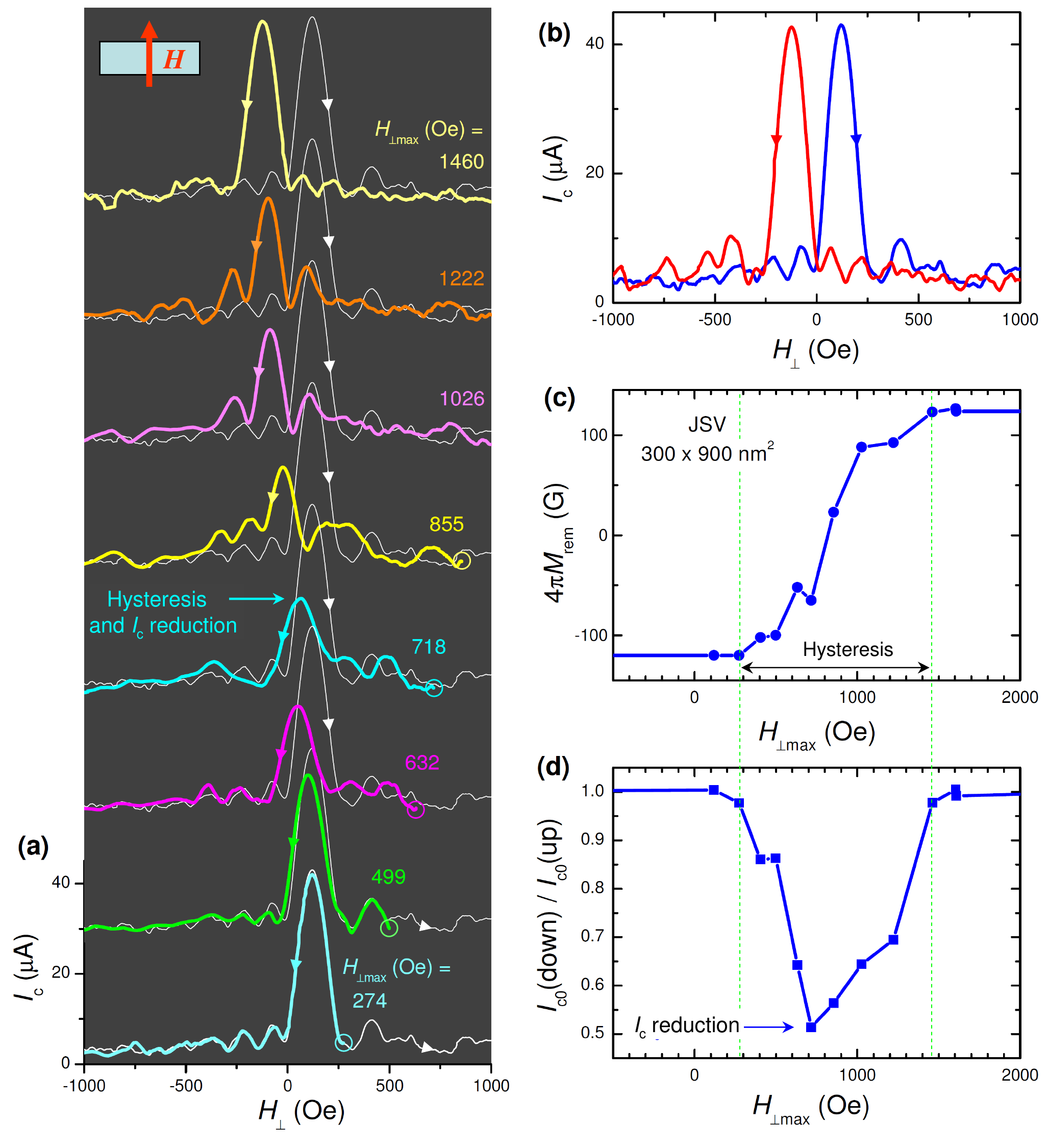}
    \caption{(color online). Experimental FORC analysis for a JSV's $300\times 900$ nm$^2$ in the hard axis orientation
    at $T=1.0$ K. (a) Thin white lines represent $I_c(H_{\perp})$ for
    the upward field sweep starting from the saturated $\downarrow\downarrow$ state. Thick color
    lines are reversal curves starting from different $H_{\perp
    max}$. The curves are shifted vertically for clarity. $H_{\perp
    max}$ are indicated by circles and/or text. (b) Mirror-symmetric $I_c(H_{\perp})$ curves for upward (blue) and downward (red) field
    sweeps from saturated $\downarrow\downarrow$ and $\uparrow\uparrow$
    states. (c) Position of the central maximum of $I_c(H_{\perp})$ FORC's as a function of the
    reversal field $H_{\perp max}$. It represents absolute values of remnant
    magnetization of the JSV. (d) Amplitudes of the central maximum of downward FORC's $I_{c0}(down)$,
    normalized by that for the upward sweep $I_{c0}(up)$ as a function of the reversal field $H_{\perp max}$.
    Note that in contrast to SFS junctions, Fig. \ref{fig:fig2} (f), remagnetization of the JSV leads to a non-trivial hysteresis, which is accompanied by a significant reduction of the supercurrent.}
    \label{fig:fig4}
\end{figure}

In Figure \ref{fig:fig5} we analyze behavior of the $160 \times 860$ nm$^2$ JSV's in the easy
axis orientation. Fig. \ref{fig:fig5} (a) represents FORC analysis.
FORC's are reversible untill $H_{\parallel max}$ passes the first $I_c(H_{\parallel})$ peak 
in the uppward sweep (thin white lines). At higher fields hysteresis appears, accompanied by the reduction of $I_c$. The $I_c$ reaches a minimum when $H_{\parallel max}$ passes the
second maximum at 816 Oe. At $H_{\parallel max}=916-1473$ Oe a state with one dominant peak is
observed. With further increase of $H_{\parallel max} \geq 1475$ Oe, the
second peak reemerges. Finally, for $H_{\parallel max}$ larger than the
saturation field, $\simeq 2$ kOe, the reversal curve becomes
mirror symmetric with respect to the uppward curve. Thus, hysteresis in JSV's is
non-trivial for both field orientations: the appearance of
hysteresis is always accompanied by the reduction of
supercurrent, as indicated in Figs. \ref{fig:fig4} (d) and
\ref{fig:fig5} (c).

\subsection{C. Difference between SFS junctions and JSV's}

To understand the difference in behavior of JJ's and JSV's, we first note that the conventional Fraunhofer
$I_c(H)$ modulation in JJ's occurs due to flux quantization with field independent critical current density, $J_c(H)=$ const.
The observed trivial hysteresis in SFS junctions suggests that upon remagnetization of a single F-layer only the total flux changes, but $J_c$ remains unchanged.
Conversely, the non-trivial hysteresis in JSV's indicates that $J_c$ is not constant, but depends on the relative orientation $\alpha$ of the two F-layers.
It is anticipated  \cite{Trifunovic_2011,Trifunovic_PRL2011,Melnikov_2012,Iovan_2017} that the triplet component should vanish in the collinear $\alpha =
0, 180, 360^{\circ}$ states and should have maxima in the noncollinear
$\alpha = 90, 270^{\circ}$ states, see numerical analysis in the Appendix.

The origins of magnetic hysteresis in JJ's and JSV's are also different. For JJ's with a single F-layer it is caused predominantly by the shape anisotropy.
Presence of the second F-layer in JSV's leads to another mechanism caused by magnetostatic interaction between F$_{1,2}$-layers, which favors the AP state. In a
mono-domain case remagnetization of a JSV starts by a scissor-like rotation of $M_{1,2}$ \cite{Iovan_2017}.
Such rotation is reversible and non-hysteretic.
It continues until the softer F$_1$-layer flips and JSV switches into the AP state. Magnetostatic
stability of the AP state leads to the appearance of hysteresis: if
the field is reversed, the spin valve will remain in the AP state.
With increasing field the harder F$_2$-layer also flips and JSV enters into the second scissor-like noncollinear state,
which gradually turns into the parallel $\uparrow\uparrow$ state \cite{Iovan_2017}.
Micromagnetic simulations for our JSV's, presented in the Appendix, confirm such
a behavior but also demonstrate that remagnetization may involve
formation of two domains. Few domains do not change the overall
picture, but lead to additional hysteresis associated with
the disappearance of each domain wall.

\begin{figure}[t]
    \centering
    \includegraphics[width=0.5\textwidth]{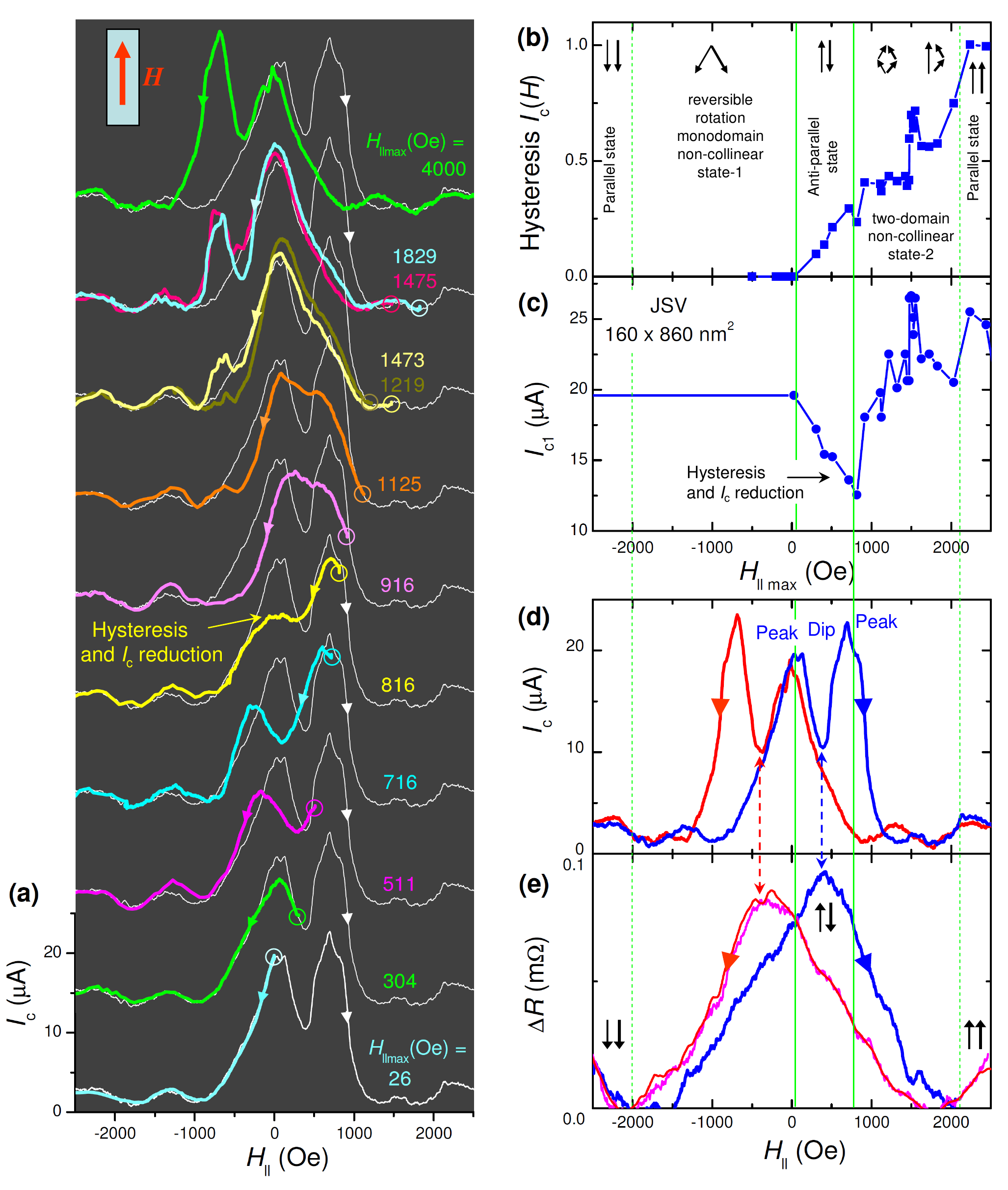}
    \caption{(color online). {\em In-situ} characterization of a JSV $160\times 860$ nm$^2$
in the easy axis orientation at $T=1.2$ K.
    (a) Experimental FORC's (thick color lines) for different $H_{\|max}$, indicated by circles. The curves are shifted vertically for
    clarity. (b) Hysteresis of $I_c(H)$, equal to the area between upward and reversal curves in
    (a). (c) Amplitude of the first supercurrent peak $I_{c1}(H\sim 0)$.
    Arrows in the top of panel (b) depict the magnetic state of the JSV, as described in the text.
    (d) $I_c(H_{\|})$ patterns for uppward (blue) and downward (red) field sweeps from saturated $\downarrow\downarrow$ and $\uparrow\uparrow$ states. (e) Spin-valve magnetoresistance, measured at
    bias current much larger than $I_c$. Dashed vertical arrows in (d,e) indicate a
    clear correlation between the dip in $I_c$ and the maximum of MR, corresponding
    to the AP state of the spin-valve.
}
    \label{fig:fig5}
\end{figure}

To summarize the above discussion, the principle difference between JJ's and JSV's is in $J_c(H)$ dependence, which is constant for JJ´s and depends on magnetization orientation, $J_c[\alpha(H)]$, for JSV's. During remagnetization $\alpha(H)$ varies from 0 to $360^{\circ}$ passing two times through noncollinear
states $\alpha = 90^{\circ}$ and $270^{\circ}$. Therefore, the triplet component should have two peaks at $\alpha = 90^{\circ}$ and $270^{\circ}$,
surrounding a dip in the AP state $\alpha =180^{\circ}$, while the singlet component should have one sharp maximum in the AP state, see the Appendix.
This provides a robust qualitative test for the nature of the dominant supercurrent: Since the appearance of hysteresis in JSV is caused by the switching from the noncolliniar scissor-state to the collinear AP state, the associated change in $I_c$ should unambiguously reveal the dominant type of supercurrent. If $I_c$ increases than
it is singlet and if decreases - triplet. The later is qualitatively consistent with our observations, see Figs. \ref{fig:fig4} (d) and \ref{fig:fig5} (c).

\subsection{D. {\em In-situ} characterization of JSV state}

Unambiguous confirmation of the triplet nature of supercurrent requires
detailed knowledge of the micromagnetic state.
Figs. \ref{fig:fig5} (b-e) represent the {\em in-situ} analysis of the
magnetic state evolution for the easy axis orientation of the JSV. Fig.
\ref{fig:fig5} (b) represents hysteresis, i.e., area
between uppward $I_c(H_{\parallel})$ and FORC's from Fig. \ref{fig:fig5} (a). Fig.
\ref{fig:fig5} (c) shows amplitudes of the first (left) main peak
$I_{c1}$ in FORC's. Fig. \ref{fig:fig5} (e) shows high-bias spin-valve magnetoresistance measured at the same device
\cite{Supplem}. Parallel and AP states of JSV correspond to minima and maxima of MR, respectively \cite{Iovan_2014}.
Such the analysis provides a self-consistent understanding of the magnetic state evolution in the JSV. The saturation
field, at which FORC's stop changing, see Fig. \ref{fig:fig5} (b),
and MR reaches minimum, see Fig. \ref{fig:fig5} (e), is $\sim \pm
2$ kOe. At $H<-2$ kOe the JSV is in the $\downarrow\downarrow$
parallel state $\alpha \simeq 0$. In a broad range -2 kOe $<H_{\parallel max}<26$ Oe, there is no hysteresis.
Consequently, the JSV is in a mono-domain noncollinear scissor state with reversible rotation $0<\alpha <180^{\circ}$.
Hysteresis appears at $H_{\parallel max} \gtrsim 26$ Oe, indicating switching
into the magnetostatically stabile AP state $\alpha \simeq 180^{\circ}$, as confirmed by the large value of MR.
At $H_{\parallel max}>816$ Oe a sudden change occurs both in hysteresis, Fig.
\ref{fig:fig5} (b), and $I_{c1}$, Fig. \ref{fig:fig5} (c). It indicates a switching out of the AP state into a second noncollinear state $180^{\circ}< \alpha <360^{\circ}$. At $H_{\parallel max}\simeq
1473$ Oe there is yet another jump in both hysteresis, and
$I_{c1}$, before reaching the saturated $\uparrow\uparrow$ parallel state, $\alpha = 360^{\circ}$, at $\sim 2$ kOe.
Such a two-step switching from AP to $\uparrow\uparrow$ parallel state is fully
consistent with micromagnetic simulations presented in the Appendix and is due to formation of two
domains in both layers. At $H_{\parallel max}\simeq 1473$ Oe the thinner
F-layer switches into the monodomain state, followed by the
thicker F-layer close to the positive saturation field.
Arrows in the top part of Fig. \ref{fig:fig5} (b) sketch the evolution of magnetic states during the remagnetization.

\subsection{E. Correlation between the supercurrent and the magnetic state in JSV}

Now we can look at correlations between the supercurrent and the
magnetic state. In Fig. \ref{fig:fig5} (d) we show $I_c(H)$
patterns for this JSV. Let's focus on the upward field sweep (blue
line). It has a double-peak structure. Solid vertical lines going
through Figs. \ref{fig:fig5} (b-e) emphasize that all {\em
in-situ} characterization methods unanimously connect the dip with
the AP state. Most straightforwardly this is seen from comparison
with the MR. Dashed arrows in Figs. \ref{fig:fig5} (d,e) indicate that
the dip in $I_c$ corresponds to the maximum of MR. Furthermore,
FORC analysis, Figs. \ref{fig:fig5} (b,c), indicates that the
field range of the primary hysteresis, $26$ Oe $<H_{\|max}
\lesssim 816$ Oe, associated with magnetostatic stability of the
AP state, coincides with the field range between the two peaks and
that the appearance of this hysteresis is accompanied by the reduction of
the $I_{c1}$ peak. Consequently, entrance into the AP state leads to a
significant reduction of $I_c$ through the JSV. However, the supercurrent recovers
upon leaving the collinear AP state in both direction, resulting
in the observed double-peak $I_c(H)$ pattern. We emphasize that
such the behavior is opposite to expectations for the singlet current,
which should be at maximum in the AP state and is consistent with the
predictions for the odd-frequency spin-triplet supercurrent, see the Appendix.

We note that such an unsusual behavior has not been reported in an earlier similar work \cite{Iovan_2014} on JSV's containing dilluted CuNi ferromagnets because in that case the dominant supercurrent ($\sim 80 \%$) had a singlet nature. An estimation based on numerical fitting of our data, presented in the Appendix, indicates that in our Ni-based JSV's the triplet current amplitude is approximately three times larger than the singlet. This helps to uncover the
characteristic double-peak modulation, which provides an unambiguous evidence for generation of the spin-triplet order parameter.
Yet, even in Ni-based JSV's the dip in the AP state does not go to zero, indicating that
there is still a significant subdominant ($\sim 30\%$) singlet supercurrent even through a relatively thick Ni.

\section{V. Conclusion}

To conclude, both singlet and triplet supercurrents can flow through S/F heterostructures
\cite{Buzdin2005,Efetov2005,Fominov,Blanter2004,Eschrig,Houzet_2007,Golubov,Trifunovic_2011,Trifunovic_PRL2011,Melnikov_2012,Pugach_2012,Linder_2012,Richard_2015,Hikino_2015,Linder_2015,Ren_2016}.
The unique feature of our work that adds to further understanding of the triplet state, along with earlier
experimental works \cite{Bell_2004,Robinson_2010,Iovan_2014,Dresselhaus_2014,Ovsyannikov,Robinson_Sc2010,Khaire_2010,Banerjee_2014,Glick_2017,Aarts_2017,Martinez_2016,Strunk_2017},
was development of {\em in-situ} characterization techniques for an
accurate assessment of micromagnetic states in actual
nano-devices. In particular, we developed a new AJF+FORC technique, a powerful tool allowing absolute magnetometry of
nano-devices and accurate identification of micromagnetic states.
We fabricated and studied nano-scale Josephson junctions and (pseudo) spin valves with Ni-interlayers. Small sizes enabled mono-(or few) domain configurations, which could be unambiguous identificatied. A strong F (Ni) was employed for
reduction of the singlet current, enabling the dominant triplet
component. This was instrumental for observation of an extraordinary behavior of JSV's, qualitatively different from similar-size SFS JJ's.
Namely, $I_c(H)$ modulation of studied JSV's had two main peaks separated by a dip and exhibited a non-trivial hysteresis, accompanied by reduction of $I_c$.
The {\em in-situ} characterization showed a clear correlation of the
$I_c$ dip with the antiparallel state of the spin valve and the two peaks to the two noncollinear states aside of it, thus providing unambiguous evidence for generation of the spin-triplet order parameter.

\appendix

\section{ Appendix: Numerical modeling of Josephson spin-valve}

\begin{figure*}[t]
    \centering
    \includegraphics[width=0.9\textwidth]{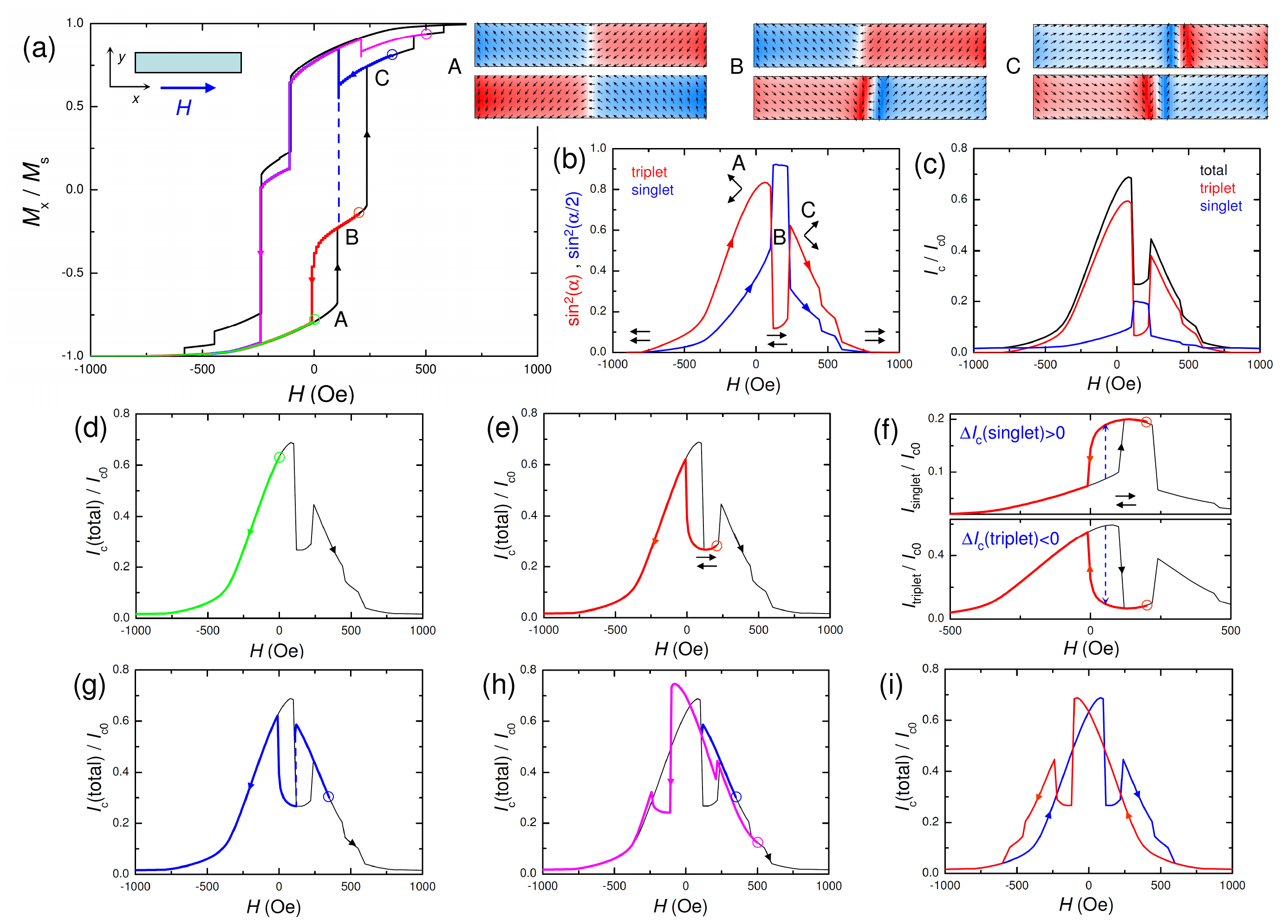}
    \caption{Simulations for easy axis field orientation. (a) Magnetization curves $M_x(H_x)$.
    Black curves represent the major hysteresis and color curves FORC's with different
    $H_{\parallel max}$, indicated by circles. The top right panel shows magnetization distribution in $F_{1,2}$ layers at points A,B
    and C. (b) Sample-averaged values of normalized triplet ($\propto \sin^2(\alpha)$, red) and singlet
    ($\propto \sin^2(\alpha/2)$, blue) current amplitudes for
    the upward field sweep. It is seen that the triplet amplitude has a minimum in the AP state-B, surrounded by the two maxima
    in the noncollinear states A and C. The singlet current has a single maximum in the AP state-B.
    (c) Corresponding critical currents for the case when triplet amplitude is $\sim 3$ times larger than the singlet.
    (d-i) Simulated FORC's for the reversal curves from (a). Note that the hysteresis appears when the JSV switches into the magnetostatically stable AP state-B, see panel (e). Panel (f) demonstrates that in this hysteresis the singlet current is enhanced, while the triplet is reduced. Therefore, enhancement/reduction of $I_c$ upon appearance of the hysteresis is an unambiguous fingerprint of dominant singlet/triplet supercurrent. }
    \label{fig:fig6}
\end{figure*}

\begin{figure*}[ht]
    \centering
    \includegraphics[width=0.9\textwidth]{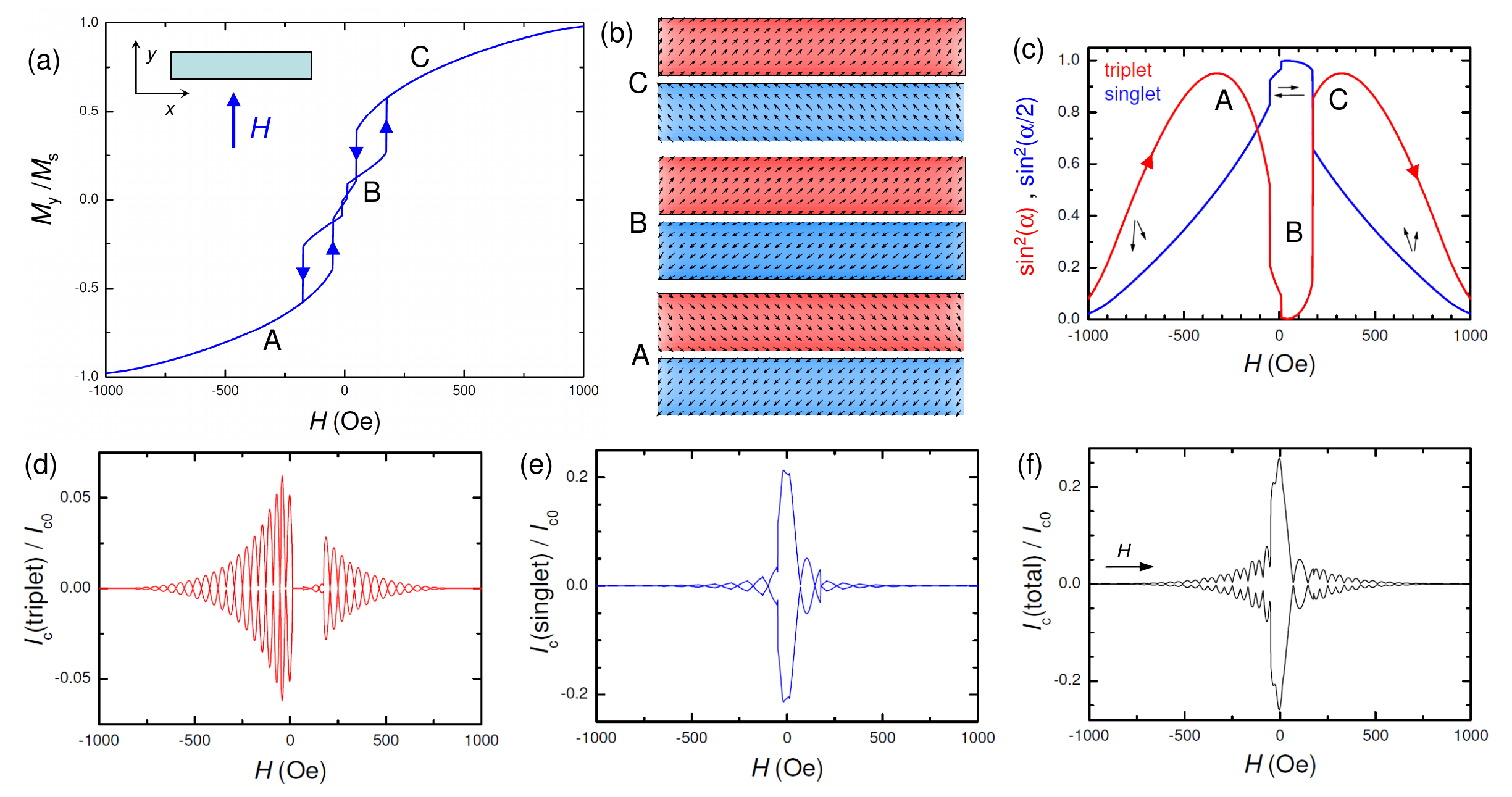}
    \caption{Simulations for the hard axis orientation.
   (a) Magnetization curves. (b) Configurations of magnetization at points A, B, C.
   (c) Sample-averaged values of normalized triplet ($\propto \sin^2(\alpha)$, red) and singlet
    ($\propto \sin^2(\alpha/2)$, blue) current amplitudes for the upward field sweep.
   Panels (d-f) show magnetic field modulation patters for (d) triplet,
    (e) singlet, and (f) total critical currents. It can be seen that although triplet current amplitude in (c) has two
    clear peaks in noncollinear states A, C, however, small flux quantization field at this
    field orientation leads to rapid damped oscillations, which suppresses the triplet supercurrent (see the vertical scale). Therefore, the characteristic double-maxima become unrecognizable in the total $I_c(H)$ modulation (f). This explains how flux quantization effect changes $I_c(H)$ patterns for JSV's from a double- to a single-peak for easy- and hard-axis orientations, respectively.
    }
    \label{fig:fig7}
\end{figure*}

To clarify the behavior of JSV we perform numerical analysis.
Josephson current in JSV's has three main components \cite{Melnikov_2012}: the
short-range spin-singlet $I_{ss}$, the long-range spin singlet
$I_{sl}$ and the long-range spin-triplet $I_{tr}$.
Their local values depend on relative angles, $\alpha(x,y)$, between
magnetizations, $M_{1,2}$, of the two F-layers and the
Josephson phase difference between S-electrodes $\varphi(x,y)$:
\begin{eqnarray}\label{Ic123}
I_{ss}(x,y) = I_{ss0} \cos^2[\alpha(x,y)/2]\sin[\varphi(x,y)],\\
I_{sl}(x,y)=I_{sl0} \sin^2[\alpha(x,y)/2]\sin[\varphi(x,y)],\\
I_{tr}(x,y) = I_{tr0} \sin^2[\alpha(x,y)]\sin[2\varphi(x,y)].
\end{eqnarray}
To calculate $I_c(H)$ we follow the procedure from Ref. \cite{Iovan_2017}. First we perform a micromagnetic simulation in OOMMF, which provides the two-dimensional distribution of components
$M_{x1,2}(x,y)$ and  $M_{y1,2}(x,y)$. Next, we calculate $\varphi$ by direct integration of:
\begin{eqnarray}\label{dFixy}
\frac{\partial\varphi(x,y)}{\partial y}=
\frac{2\pi d^*}{\Phi_0}H_x+\frac{2\pi d_1}{\Phi_0}4\pi M_{x1}+\frac{2\pi d_2}{\Phi_0}4\pi M_{x2},~~\\
\frac{\partial\varphi(x,y)}{\partial x}= \frac{2\pi
d_1}{\Phi_0}4\pi M_{y1}+\frac{2\pi d_2}{\Phi_0}4\pi M_{y2}.~~~~~
\end{eqnarray}
Here $H_x$ is the applied magnetic field in the $x$-direction and $d_{1,2}$ are the thicknesses of F$_{1,2}$ layers.
The total supercurrent $I_s=I_{ss}+I_{sl}+I_{tr}$, Eq. (1-3), is
integrated over the JSV area using the obtained values
$\alpha(x,y)$ and $\varphi(x,y)$. To find the critical current we
maximized the supercurrent with respect to the integration
constant. For more details of the simulation procedure see Ref. \cite{Iovan_2017}.
In Figures \ref{fig:fig6} and \ref{fig:fig7} we show results corresponding to one of the studied JSV's Ni(5nm)/Cu(10nm)/Ni(7.5nm) with sizes $160 \times 860$ nm$^2$. Simulations are shown for the following set of supercurrent amplitudes: $I_{ss0}=0.1$, $I_{sl0}=1.0$, $I_{tr0}=3.0$, which fits well the experimental data. From this we conclude that the triplet current amplitude in our JSV's is approximately three times larger than the singlet, $I_{tr0}/(I_{sl0}+I_{ss0})\simeq 3$ .

Fig. \ref{fig:fig6} (a) shows the magnetization curve along the easy
axis (see the inset). Black lines represent the major hysteresis
loop and color lines - FORC's with $H_{\parallel max}$ indicated by circles.
The spin valve appears to be at the border between the mono- and
the two-domain states. Upon sweeping of the field upwards from the saturated
$\downarrow\downarrow$ state, magnetization in F-layers first
curves into a C-shape (state-A), which is reversible without hysteresis (see the green line).
Then the $F_1$ and $F_2$ layers switch sequentially into the state with two domains (states B and
C) simultaneously flipping the $x$-component of magnetization. Hysteresis appears in the state B (red line), which corresponds to the AP state.

Fig. \ref{fig:fig6} (b) shows amplitudes of the long-range singlet
(blue) and triplet (red) supercurrents for an upward field sweep.
In the AP state-B the singlet amplitude is large and the triplet is small. On both sides of it, there are two noncollinear states
A, C with large triplet and small singlet amplitudes. At large
positive/negative fields the JSV is in the parallel state with vanishing of both singlet and triplet
long-range components. The shape of $I_c(H)$ pattern of the JSV depends on relative
amplitudes of singlet and triplet components. Fig.
\ref{fig:fig6} (c) shows the case with the dominant triplet
current ($I_{ss0}=0.1$, $I_{sl0}=1.0$, $I_{tr0}=3.0$) for the total
(black), singlet (blue) and triplet (red) currents. Since in this case
the total current is dominated by the triplet current, $I_c(H)$  has two peaks corresponding to the
noncollinear states A and C,
separated by a dip, corresponding to the AP state B, similar to the experimental data in Fig. \ref{fig:fig5} (d).

In Figs. \ref{fig:fig6} (d-i) we analyze $I_c(H_{\parallel})$ FORC's, corresponding to $M(H)$ curves
with the same color in Fig. \ref{fig:fig6} (a). Panel (d)
represents the case when $H_{\parallel max}$ is within the first $I_c(H)$
peak. Here the spin valve is in the reversible noncollinear state-A. In (e) $H_{\parallel max}$ is within
the dip in the AP state-B. As emphasized in the main text, the fingerprint of the AP state is the appearance of hysteresis,  see red line in Fig.
\ref{fig:fig6} (a). Panel (f) demonstrates that within this hysteresis the singlet current is
increased (top panel) and the triplet is decreased (bottom panel). Thus, the change of the current upon appearance of hysteresis
tells us about the nature of the dominant supercurrent. Since in our simulations the triplet current is dominant, there is an overall
drop of $I_c$ at the hysteretic branch, as seen in Fig. \ref{fig:fig6} (e).
Panels (g) and (h) show FORC's after switching out of the AP state B into the noncollinear state C with domains. Note that along with some metastability
associated with domains, in Fig. \ref{fig:fig6} (h) we observe a net enhancement of the central noncollinear peak at the expense of
the second peak. Finally, panel (i) shows $I_c(H)$ starting from
fully saturated states. Overall, presented simulations are in a good agreement with
experimental data for JSVs' from Fig. \ref{fig:fig5} (a). Simulations reproduce both the double-peak
$I_c(H_{\parallel})$ patterns and the nontrivial hysteresis with reduction of $I_c$ in the AP state.

We note that we assumed that the JSV is narrow enough so that flux
quantization field is larger than the saturation field. Therefore,
critical current modulation is not upset by flux quantization.
However, in larger JSV's flux quantization does strongly affect the
$I_c(H)$ modulation \cite{Iovan_2017}.
This is the main reason for size-dependence of $I_c(H)$ patterns, see Fig. \ref{fig:fig3}.
For long JSV's with a small $\Delta H$ the double-peak structure of $I_{tr0}(H)$
is completely masked by rapid flux-quantum oscillations, leading
an overall Fraunhofer-type modulation of $I_c(H)$.

To demonstrate this, in Figure \ref{fig:fig7} we present simulations for
the same device in the hard axis orientation with larger $L=860$ nm, see the sketch in Fig. \ref{fig:fig7} (a).
Fig. \ref{fig:fig7} (a) shows the large hysteresis of magnetization curve $M_y(H_y)$. Here the
intermediate AP step is also present, but with a limited range of
existence, compared to the easy axis, Fig. \ref{fig:fig6} (a). This occurs
because at $H=0$ moments tend to align with the easy $x$-axis,
destroying the AP state. To the contrary, the range of fields for
coherent rotation of magnetization is broader and both layers
remain in the monodomain state. Corresponding distributions of
magnetizations are shown in Fig. \ref{fig:fig7} (b) for
points A) the first non-collinear state upon coherent rotation
from the negative parallel state, B) antiparallel state and C) the
second non-collinear state upon switching from the AP state.

Fig. \ref{fig:fig7} (c) shows sample-averaged values of
normalized triplet, $\propto \sin^2(\alpha)$ (red) and singlet,
$\propto \sin^2(\alpha/2)$, (blue) current amplitudes, see Eqs.
(3) and (2). The behavior of both components is similar to the easy axis case, Fig. \ref{fig:fig6} (b). I.e., in this respect the field orientation
does not make a principle difference. However, the $I_c(H)$ pattern is strongly affected in this orientation.
Figs. \ref{fig:fig7} (d-f) show magnetic field modulation of (d) triplet, (e) singlet, and (f) total
currents. It can be seen that although triplet current amplitude
in panel (c) has two clear peaks (A, C), however, the large length $L$ of the JSV at this field orientation
makes $\Delta H$ much smaller than the coercive field.
Therefore at points A and C with the largest triplet amplitudes there are already many flux quanta
inside the JSV, suppressing the triplet critical current by more than an orders of magnitude. As a result, the
characteristic double-maxima feature becomes unrecognizable in the
total $I_c(H)$ modulation. Thus, the difference between easy and hard axis orientations is entirely due to the flux quantization effect.
Nevertheless, both numerical simulations, see Figs. \ref{fig:fig6} (b) and \ref{fig:fig7} (c) and
experimental analysis in Figs. \ref{fig:fig4} and \ref{fig:fig5} demonstrate that the
essential physics remains independent of the field orientation.

\subsection{Acknowledgments}
The work was supported by the European Union H2020-WIDESPREAD-05-2017-Twinning project ``SPINTECH" under grant agreement Nr. 810144 and the Russian Science
Foundation grant No. 19-19-00594. The manuscript was written during a sabbatical semester of V.M.K.
at MIPT, supported by the Faculty of Natural Sciences at SU and
the Russian Ministry of Education and Science within the program ``5top100".



\end{document}